\documentclass[%
 aip,
 amsmath,amssymb, 
 reprint,%
 citeautoscript
]{revtex4-2}

\usepackage{hyperref}
\usepackage{graphicx}
\usepackage{bm}
\usepackage{orcidlink}

\usepackage[utf8]{inputenc}
\usepackage[T1]{fontenc}

\usepackage[cal=boondox]{mathalfa} 

\makeatletter
\def\@email#1#2{%
 \endgroup
 \patchcmd{\titleblock@produce}
  {\frontmatter@RRAPformat}
  {\frontmatter@RRAPformat{\produce@RRAP{*#1\href{mailto:#2}{#2}}}\frontmatter@RRAPformat}
  {}{}
}%
\makeatother

\def\DD{\boldsymbol{\mathcal{D}}}

\def\gg{\mathbf{g}}
\def\GG{\mathbf{G}}
\def\jj{\mathbf{j}}
\def\JJ{\mathbf{J}}
\def\ff{\bm{\varphi}}

\def\rr{\mathbf{r}}
\def\RR{\mathbf{R}}
\def\vv{\mathbf{v}}
\def\VV{\mathbf{V}}

\def\half{\frac{1}{2}}

\usepackage[normalem]{ulem}
\usepackage{xcolor}
\definecolor{amber}{rgb}{1,0.49,0}
\definecolor{darkgreen}{rgb}{0,0.55,0}
\definecolor{tangerine}{rgb}{0.944,0.522,0}
\definecolor{verde}{rgb}{0.,0.6,0}
\definecolor{rosso}{rgb}{0.9,0.0,0.2}
\definecolor{magenta}{rgb}{0.9,0.2,0.9}

\newif\ifhighlight

\newcommand{\highlight}{\highlighttrue}
\highlight
\newcommand{\editor}[2]{%
  \expandafter\newcommand\csname #1note\endcsname[1]{%
    \textcolor{#2}{(\textbf{#1:} ##1)}}%
  \expandafter\newcommand\csname #1\endcsname[1]{%
    \ifhighlight\textcolor{#2}{##1} \else ##1\fi}%
  \expandafter\newcommand\csname #1cancel\endcsname[1]{%
    \ifhighlight\textcolor{#2}{\sout{##1}}\fi}%
  \expandafter\newcommand\csname #1change\endcsname[2]{%
    \ifhighlight\textcolor{#2}{\sout{##1} ##2}\else ##2\fi}%
  \newenvironment{#1text}{\ifhighlight\color{#2}\fi}{\color{black}}
}

\editor{RS}{cyan}

\usepackage[most]{tcolorbox} 
\tcbset{enhanced}

\newtcolorbox{strikebox}{
  enhanced,
  frame empty,        
  colback=white,      
  coltext=cyan,
  boxsep=0pt, left=0pt, right=0pt, top=0pt, bottom=0pt,
  overlay={
    \draw[line width=1.5pt, red] (frame.south west) -- (frame.north east);
    \draw[line width=1.5pt, red] (frame.north west) -- (frame.south east);
  }
}

\begin{document}
\title{The nuts and bolts of gauge invariance of heat transport}
\author{Stefano Baroni\,\orcidlink{0000-0002-3508-6663}}
\affiliation{
SISSA -- Scuola Internazionale Superiore di Studi Avanzati, Trieste \\
CNR -- IOM, Istituto dell'Officine dei Materiali, SISSA, Trieste
}

\begin{abstract}
    In this work I revisit the notion of \emph{gauge invariance} in thermal transport and show, in the simplest and most general possible terms, why heat conductivity is unaffected by the specific choice of energy density. I provide the minimal and general conditions under which any two energy densities, though differing locally, lead to the same heat conductivity within the Green–Kubo framework. The relevance of gauge invariance in heat-transport simulations performed with machine-trained neural-network potentials is also briefly highlighted.
\end{abstract}
\date{28 Septermber 2025} \maketitle

Transport coefficients quantify how the flux of conserved quantities such as energy, charge, or mass responds to small perturbations, thus linking microscopic dynamics to macroscopic irreversibility. Until about a decade ago, the adiabatic thermal conductivity of electronically gapped materials was deemed beyond the reach of quantum-mechanical first-principles calculations, because the microscopic energy density---from which the heat current is defined through the continuity equation---is intrinsically ill-defined \cite{Marcolongo2016}. Many different forms of this density are compatible with the same total energy and atomic forces, and this apparent indeterminacy seemed to undermine the very definition of heat conductivity in atomistic simulations. The effort to resolve this difficulty led to the realization that transport coefficients are not affected by it: different choices of the energy density produce different fluxes and correlation functions, but the Green–Kubo \cite{Green1952,*Green1954,Kubo1957a,*Kubo1957b} integrals that define the conductivity do not depend on them. This property, later referred to as \emph{gauge invariance of transport coefficients}, has been supported by theoretical arguments and numerical tests \cite{Ercole2016,Grasselli2021}. A minimal and general proof of its validity, however, has so far been missing. The purpose of this work is to provide such a proof, thereby placing gauge invariance on firm theoretical ground and clarifying its scope in the theory of thermal transport.

\section{Preamble}\label{sec:preamble}
According to the Green--Kubo theory of linear response \cite{Green1952,*Green1954,Kubo1957a,*Kubo1957b}, the thermal conductivity of a classical system is determined by the time integral of the autocorrelation function of the \emph{energy flux}, $\JJ_E$, 
$\kappa=\frac{k_B\beta^2}{3V} \Lambda_{EE}$, where the energy-energy diagonal element of the Onsager matrix \cite{Onsager1931} is defined as:  
\begin{align}
    \Lambda_{EE} &= \int_{0}^{\infty} \langle \JJ_E(t) \cdot \JJ_E(0) \rangle \, dt, \label{eq:green-kubo} \\
           &= \lim_{T\to\infty} \frac{1}{2T} \left\langle \left | \int_0^T\JJ_E(t)dt \right |^2 \right \rangle, \label{eq:einstein-helfand}
\end{align}
$V$ being the system’s volume, $\beta$ its inverse temperature in energy units, and $k_B$ Boltzmann’s constant. This is only true for systems—such as solids or one-component fluids—where energy is the only diffusive conserved quantity. For simplicity, I will tacitly assume this to be the case until gauge invariance is discussed and a more general multi-component approach \cite{Bertossa2019,Grasselli2021} becomes necessary. The second expression, Eq. \eqref{eq:einstein-helfand}, generalizes to the case of heat transport Einstein's relation between molecular diffusivity and the velocity auto-correlation function \cite{Einstein1905}, as reformulated by Helfand \cite{Helfand1960}. The flux $\JJ_E$ is defined as the volume integral of the microscopic \emph{energy current density} $\jj_E(\mathbf{r})$, $\JJ_E = \int_V \jj_E(\rr) \, d\rr$. Both $\JJ_E$ and $\jj_E$ depend on time through their adiabatic dependence on atomic positions, $\mathbf{R} = \{\rr_1, \rr_2, \dots, \rr_N\}$, and velocities $\mathbf{V} = \{\vv_I\}$, while the latter satisfies the continuity equation, 
\begin{equation}
    \nabla \cdot \jj_E = -\dot{\epsilon}, \label{eq:continuity}
\end{equation}
where $\epsilon(\mathbf{r}|\mathbf{R},\mathbf{V})$ is a suitably defined \emph{energy density},
\begin{equation}
    \epsilon(\rr|\RR,\VV) = \half \sum_I M_I\vv_I^2\delta(\rr-\rr_I) + w(\rr|\RR), \label{eq:energy-density}
\end{equation}
$\{M_I\}$ are atomic masses, and $w(\rr|\RR)$ is a \emph{potential-energy density} that does not depend on velocities. The expression in Eq. \eqref{eq:energy-density} also applies when the potential energy is decomposed into localized atomic contributions, $e_I$. In that case, the potential-energy density would read: $w(\rr|\RR) = \sum_I\delta(\rr-\rr_I) e_I(\RR)$.  The notation $\epsilon(\rr|\RR,\VV)$ is meant to indicate a parametric dependence of a function of $\rr$ upon whatever variables follow the vertical bar. For future reference, also note that lower boldfaces, $\rr_I$ and $\vv_I$, denote positions and velocities of individual atoms $\bigl (\in\mathbb{R}^3 \bigr )$, while their upper cases indicate collections of atomic positions and velocities $\bigl (\in\mathbb{R}^{3N} \bigr )$.

The total energy of the system is the volume integral of the density:
\begin{equation}
    \begin{aligned}
        E(\RR,\VV) &= \int_V \epsilon(\rr|\RR,\VV) d\rr \\
        &=\half\sum_IM_I\vv_I^2 + \int_V w(\rr|\RR)d\rr,
    \end{aligned}
\end{equation}
whereas atomic forces, $\mathbf{f}_I=-\frac{\partial E}{\partial\rr_I}$ can be seen as integrals of the \emph{force densities}, $\ff_I(\rr|\RR)$:
\begin{align}
    \mathbf{f}_I(\RR) &= \int_V  \ff_I(\rr|\RR) d\rr, \label{eq:total-forces} \\
    \ff_I(\rr|\RR) &= -\frac{\partial w(\rr|\RR)}{\partial \rr_I} \label{eq:force-densities}
\end{align}

In a finite system, multiplying Eq. \eqref{eq:continuity} by $\rr$ and integrating by parts over the entire volume gives:
\begin{equation}
    \begin{aligned}
        \JJ(\RR,\VV) &\doteq \int_V \jj(\rr|\RR,\VV)d\rr\\
        &= \int_V \dot\epsilon(\rr|\RR,\VV)\rr d\rr.
        \label{eq:Jdoteps}
    \end{aligned}
\end{equation}
Here and in the following the ``$E$'' subscript is dropped for clarity, unless strictly needed. In an extended system, or in a finite one 
with periodic boundary conditions (PBC), the second equation above is ill-defined because a non-decaying function has no meaningful first moment—just as a periodic charge-density 
distribution 
does not, which is why the macroscopic polarization of an insulator is not uniquely defined. \cite{Resta2007} By using Eq. \eqref{eq:energy-density} and following a procedure similar to that reported in Appendix A of Ref. \onlinecite{Ercole2016}, the energy flux can be cast into the form:
\begin{equation}
    \JJ(\RR,\VV) = \sum_I \left ( \half M_I \vv_I^2 + \DD_I(\RR) \right ) \cdot \vv_I , \label{eq:flux}
\end{equation}
where
\begin{equation}
    \DD_I(\RR) = -\int_V (\rr-\rr_I)\otimes \ff_I(\rr|\RR) \, d\rr \label{eq:Dl}
\end{equation}
and $\otimes$ denotes the tensor (outer) product of vectors, defined by $(\mathbf a \otimes \mathbf b)_{\alpha\beta} = a_\alpha b_\beta$. If the first moments of the atomic force densities defined in Eq. \eqref{eq:Dl} are finite, the flux in Eq. \eqref{eq:flux} is well defined in PBC as well. A derivation of Eqs. (\ref{eq:flux}-\ref{eq:Dl}) is presented in Appendix \ref{app:J-DI}. 

\section{Gauge invariance} \label{sec:GaugeInvariance}
The potential-energy density $w$ is intrinsically ill-defined, since all functions that yield the same forces according to Eqs.~(\ref{eq:force-densities}, \ref{eq:total-forces}) should be regarded as physically equivalent. A natural question then arises: do any two energy densities, equivalent in this sense, also yield the same heat conductivities? Let $w_1(\rr|\RR)$ and $w_2(\rr|\RR)$ be two such densities, and define their difference as $w' = w_2 - w_1$, while $\ff'_I(\rr|\RR) = -\partial w'(\rr|\RR)/\partial \rr_I$ denotes the corresponding difference in force densities. The condition that the two densities give rise to the same atomic forces can be expressed as
\begin{equation}
    \int_V \ff'_I(\mathbf{r}|\mathbf{R}) \, d\mathbf{r} = 0. \label{eq:F1=F2}
\end{equation}
It is not clear whether Eq.~\eqref{eq:F1=F2} alone is sufficient to guarantee equality of the heat conductivities obtained from the two energy densities. Although atomic forces derived from different local representations of the energy are expected to yield identical physical observables, the mathematical mechanism ensuring this identity for heat transport is not yet fully understood. In what follows, I show that this is indeed the case, at least in the physically sensible situation where the energy densities are \emph{short-sighted}, meaning that they generate short-ranged force densities:
\begin{equation} 
    \ff'_I(\rr|\RR) = 0 \quad \text{for } |\rr-\rr_I| > R_c. \label{eq:short-sightedness}
\end{equation}

In a series of recent papers \cite{Marcolongo2016,Ercole2016,Grasselli2021} it was shown that a sufficient condition for two energy fluxes, $\JJ_1$ and $\JJ_2$, to yield the same heat conductivity is that the time integral of their difference, $\JJ'=\JJ_2-\JJ_1$, over a molecular trajectory is bounded in phase space:
\begin{equation}
    \left | \int_0^T \JJ'(\RR^{t},\VV^{t}) dt \right | < K. \label{eq:boundedness}
\end{equation}
The time integral in Eq. \eqref{eq:boundedness} is often referred to as a \emph{Helfand's moment} after E. Helfand who first introduced it to compute transport coefficients.\cite{Helfand1960} We will see shortly that, in order to achieve full generality, this condition has to be slightly relaxed. 

Since the two densities only differ by their potential-energy term, the difference between the fluxes is given by:
\begin{equation}
    \begin{aligned}
        \JJ'(\RR) &= \int_V \dot w'(\rr|\RR)\, \rr\, d\rr \\
        &= \sum_I  \DD'_I(\RR) \cdot\vv_I  ,
    \end{aligned} \label{eq:J'}
\end{equation}
where
\begin{equation} 
    \begin{aligned}
        \DD'_I(\RR)&= \int_V \rr\otimes\frac{\partial w'(\rr|\RR)}{\partial\rr_I}\,d\rr,\\
                   &= -\int_V \rr\otimes \ff'_I(\rr|\RR)\,d\rr.
    \end{aligned} \label{eq:ff'moment}
\end{equation}
{A derivation of Eqs. (\ref{eq:J'}-\ref{eq:ff'moment}) is presented in Appendix \ref{app:J-DI}.} Note that the condition that the integral of $\ff'_I$ vanishes, Eq.~\eqref{eq:F1=F2}, makes $\DD'_I$ well defined, independent of the choice of origin. 
%
%

\begin{figure}[t!]
    \centering
    \includegraphics[width=0.9\columnwidth]{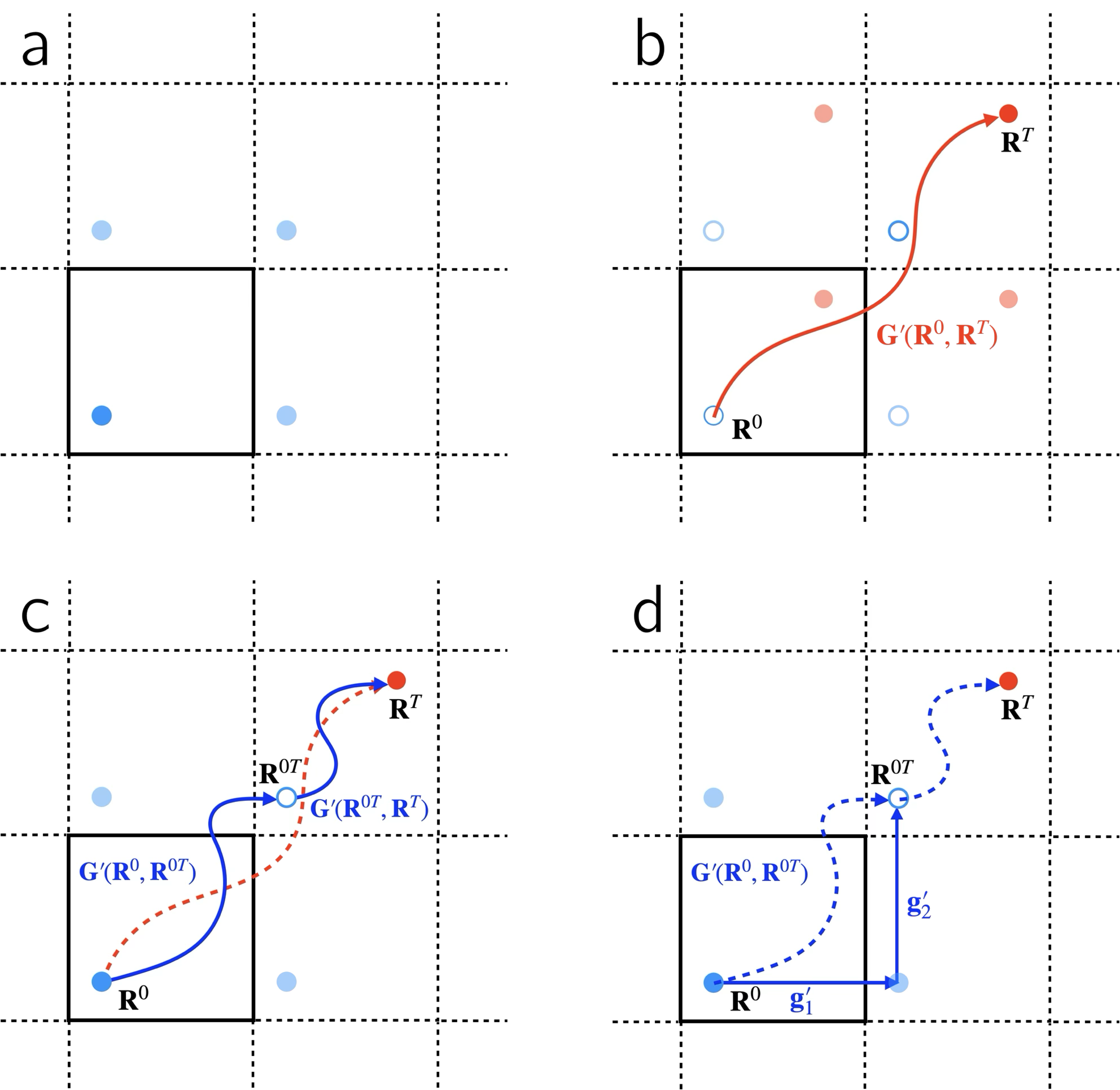}
    \caption{a) Schematic 2D representation of the $3N$-dimensional atomic configuration space with periodic boudary conditions. The blue dot represents a configuration and the pale blue dots its periodic images. b) Trajectory joining the configuration $\RR^0$ at $T=0$ to a configuration $\RR^T$ at a later time, lying in a different unit cell; $\GG(\RR^0,\RR^T)$ indicates the Helfand's moment defined in Eq. \eqref{eq:line-integral}. The physical trajectory is split into two segments, the first joining the initial configuration to its periodic image in the same unit cell as the final configuration, $\RR^{0T}$, the second from this intermediate configuration to the final one. d)  The first segment is further split into several contributions where each atom (just one in this cartoon) is displaced by a specific lattice-basis vector. Note that the ability of splitting a physical trajectory into segments of arbitary shapes is due to the exactness of the differential form, Eq. \eqref{eq:omega}. } \label{fig:Fig0}
\end{figure}

The vector-valued differential form
\begin{equation}
    \bm{\omega}(\RR) = \sum_I \DD'_I (\RR) \cdot d\rr_I \label{eq:omega}
\end{equation}
is exact, since the cross partial derivatives of its components are symmetric:
\begin{equation}
    \begin{aligned} 
        \frac{\partial \mathcal{D}'_{I\alpha\beta}(\mathbf{R})}{\partial r_{J\gamma}} &=\int_V r_\alpha \frac{\partial^2w'(\mathbf{r}|\mathbf{R})}{\partial r_{I\beta}\partial r_{J\gamma}}\,d\rr \\  &=\frac{\partial \mathcal{D}'_{J\alpha\gamma}(\mathbf{R})}{\partial r_{I\beta}}.
    \end{aligned} \label{eq:conservative-tensor-field}
\end{equation}
Therefore, the Helfand's moment in Eq.~\eqref{eq:boundedness} can be expressed as the line integral of the $\bm{\omega}$ form,
\begin{equation}
    \begin{aligned}
    \int_{0}^{T} \JJ'(\RR^{t},\VV^{t})\,dt
    &= \int_{\RR^0}^{\RR^{T}} \sum_{I} \DD'_I(\RR) \cdot d\rr_I,  \\
    &\doteq \GG'(\RR^0,\RR^T),
    \end{aligned}
    \label{eq:line-integral}
\end{equation}
taken along \emph{any} path joining the configurations at $t=0$ ($\RR^0$) and $t=T$ ($\RR^{T}$) and thus only depending on its extrema. When using PBC, the upper limit of Eq. \eqref{eq:line-integral} must be understood as the \emph{unfolded} value of the velocity integral: $\RR^T=\RR^0+\int_0^T \VV^tdt$, as illustrated in Fig. \ref{fig:Fig0}b. For Eq.~\eqref{eq:conservative-tensor-field} to hold, differentiation and integration must commute; this, in turn, requires that the integral in Eq.~\eqref{eq:F1=F2} be absolutely convergent, a condition ensured by the short-rangedness of the integrand, Eq. \eqref{eq:short-sightedness}.

Exploiting the exactness of the $\bm{\omega}$ differential form, Eq. \eqref{eq:omega}, and following the argument used for the topological definition of oxidation numbers~\cite{Grasselli2019}, the long-time behavior of the Helfand's moment can be analyzed by splitting it into two contributions,
\begin{equation}
    \GG'(\RR^0,\RR^T)
            =  \GG'(\RR^0,\RR^{0T}) + \GG'(\RR^{0T},\RR^T), \label{eq:G-def}
\end{equation}
where $\RR^{0T}$ is the periodic image of $\RR^0$ in the unit cell of $\RR^T$, expressed by
\begin{equation}
    \RR^{0T} = \RR^0 + \sum_{I\alpha} n_{I\alpha} \hat\RR_{I\alpha},
\end{equation}
with integers $n_{I\alpha}$ and displacement vectors $\hat\RR_{I\alpha}\in\mathbb{R}^{3N}$, defined as configuration differences whose components all vanish except the three coordinates of atom $I$, which equal those of the $\alpha$-th lattice-basis vector, $\bm{\tau}_\alpha$: i.e. the $J$-th atomic component of the $\hat\RR_{I\alpha}$ configuration reads $\bigl (\hat\RR_{I\alpha} \bigr )_J=\delta_{IJ}\bm{\tau}_\alpha$, see Fig. \ref{fig:Fig0}c. 
 $\GG'(\RR^0,\RR^{0T})$ can be computed by further splitting the integration path into segments where one atom at a time is displaced by a specific lattice-basis vector, $\bm{\tau}_\alpha$:
\begin{equation}
    \GG'(\RR^0,\RR^{0T}) = \sum_{I\alpha}n_{I\alpha} \gg'_{I\alpha}(\RR^0), \label{eq:g-def}
\end{equation}
where
\begin{equation}
    \begin{aligned} 
        \\
        \gg'_{I\alpha}(\RR^0) &=\int_{\RR^0}^{\RR^0+\hat\RR_{I\alpha}} \sum_J\DD'_J(\RR)\cdot d\rr_J \\
                             &= \int_{\rr^0_I}^{\rr^0_I+\bm{\tau}_\alpha} \DD_I'(\RR)\cdot d\rr_I.
    \end{aligned} 
\end{equation}
See Fig. \ref{fig:Fig0}c. 
$\gg'_{I\alpha}(\RR^0)$ does not depend on $\RR^0$, as $\frac{\partial\gg'_{I\alpha}}{\partial\RR^0} = \DD'_I(\RR^0+\hat\RR_{I\alpha}) - \DD'_I(\RR^0)=0$, because of periodicity of $\DD'_I$. The independence of $\gg'_{I\alpha}$ on $\RR^0$ is further discussed in Appendix \ref{app:Theorem}. Assuming isotropy, $\gg'_{I\alpha}$ has to be proportional to the $\alpha$-th lattice-basis vector: $\gg'_{I\alpha}=q_I\bm{\tau}_\alpha$. 
Eq. \eqref{eq:g-def} can therefore be written as:
\begin{equation}
\begin{aligned}
    \GG'(\RR^0,&\RR^{0T}) = \sum_{I\alpha} q_I n_{I\alpha}{\bm \tau}_\alpha = \sum_I q_I(\rr^{0T}_I-\rr^0_I) \\
                         &=\sum_I q_I (\rr^T_I-\rr^0_I) -\sum_I q_I(\rr^T_I-\rr^{0T}_I) \\
                         &= \sum_I \int_0^T q_I \vv_I^t dt - \sum_I q_I(\rr^{T}-\rr^{0T}_I). 
\end{aligned}
\end{equation}
The Helfand's moment in Eq. \eqref{eq:boundedness} can thus be cast as
\begin{multline}
    \int_0^T \JJ'(t)dt = \sum_{I} \int_0^T q_I \vv^t_I dt \\ -\sum_{I} q_{I}(\rr^T_I-\rr^{0T}_I) + \GG'(\RR^{0T},\RR^T). \label{eq:Jsplit}
\end{multline}
The first term in Eq. \eqref{eq:Jsplit} is a \emph{convective flux} (i.e. a linear combination of mass fluxes), whereas the last two are bounded by the size of the unit cell and can therefore be neglected in the large-time limit we are interested in. 

We conclude that energy fluxes obtained from two different local, short-sighted representations of the energy that yield the same atomic forces (two \emph{energy gauges}) can only differ by a convective flux, plus other non-diffusive terms---whose time integral stays bounded---that do not contribute to the transport coefficient. In a one-component system—such as a solid or a single-component fluid—energy is the sole diffusing conserved quantity, and the only convective flux is the total number of particles, which is proportional to the total momentum. As momentum is conserved, this flux can be disregarded when evaluating transport coefficients, and the integral of the flux differences is bounded as in Eq.~\eqref{eq:boundedness}, thus ensuring gauge invariance. In multi-component systems (such as e.g. a super-ionic solid or multi-component fluid) a full multivariate analysis is necessary.\nobreak

For the sake of definiteness, let's consider a two-component fluid, as in Ref. \onlinecite{Bertossa2019}. The relevant diffusing fluxes are energy and the number of particles of one of the two components, $\JJ_E$ and $\JJ_N$, the second number flux being constrained by momentum conservation. Energy and number fluctuations are coupled at hydrodynamic time scales, resulting in the Onsager  linear relations between fluxes and thermodynamic forces:\cite{Onsager1931}
\begin{equation}
    \begin{aligned}
        \JJ_E &= \Lambda_{EE}\nabla\left ( \frac{1}{T} \right ) - \Lambda_{EN} \nabla \left ( \frac{\mu}{T} \right ) \\
        \JJ_N &=  \Lambda_{NE} \nabla \left ( \frac{1}{T} \right ) - \Lambda_{NN} \nabla \left ( \frac{\mu}{T} \right ),
    \end{aligned} \label{eq:Onsager}
\end{equation}
where $\Lambda_{ij}=\frac{1}{3Vk_B}\int_0^\infty \left\langle \JJ_i(t)\cdot\JJ_j(0) \right\rangle dt$, $i,j\in \{E,N\}$ and $\mu$ is the chemical potential of the molecular species being considered. The thermal conductivity is defined as the ratio between the energy flux and temperature gradient, \emph{when all the number fluxes vanish}. By inserting this condition into Eqs. \eqref{eq:Onsager}, we get:
\begin{equation}
    \kappa = \Lambda_{EE} - (\Lambda_{EN})^2/\Lambda_{NN}.
\end{equation}

This expression, technically known as \emph{the Schur complement of the convective block in the Onsager matrix},\cite{Bertossa2019} is invariant under the transformation
$\mathbf{J}_E \to \mathbf{J}_E + c\,\mathbf{J}_N$.
More generally, in Ref. \onlinecite{Bertossa2019} it is shown that two energy fluxes whose difference is purely convective yield the same heat conductivity. As a welcome side consequence, although heat conductivity is defined via the heat flux $\mathbf{J}_Q=\mathbf{J}_E-\sum_s h_s\,\mathbf{J}_{N_s}$, where $\JJ_{N_s}$ is the number flux of the $s$-th molecular species and $h_s$ its partial enthalpy, it can be computed directly from the energy flux—as implicitly assumed so far—thus avoiding any cumbersome calculation of partial enthalpies.\cite{Debenedetti1988} Building on our previous result that different local, short-sighted, energy representations producing the same forces can only differ by a convective term, these observations lead to the final formulation of gauge invariance: \emph{any two short-sighted local representations of the system’s energy that yield the same atomic forces also yield the same heat conductivity}, as expected on purely physical grounds.

\section{Conclusions}

The results presented here apply not only to the representation of energy via a local density, but also to discrete representations, such as its partition into atomic contributions, which can be cast as a particular case of the former by the way of delta functions. My goal was to place  gauge invariance of the heat conductivity on a theoretically strong and general footing. Earlier sufficient criteria remain valid---notably, the boundedness in phase space of the difference between the energy fluxes associated with two different gauges. We have seen, however, that this condition can and should be released: two gauges for the local energy—continuous-density or atomistic—are equivalent whenever their energy fluxes differ by a purely convective current, i.e., a linear combination of mass currents. In this perspective, boundedness is a special sufficient case encompassed by convective invariance. I expect similar results to hold in the presence of long-range forces as well, although a detailed analysis of that case is left for future work.

\begin{figure}[ht!]
\centering
\includegraphics[width=0.95\linewidth]{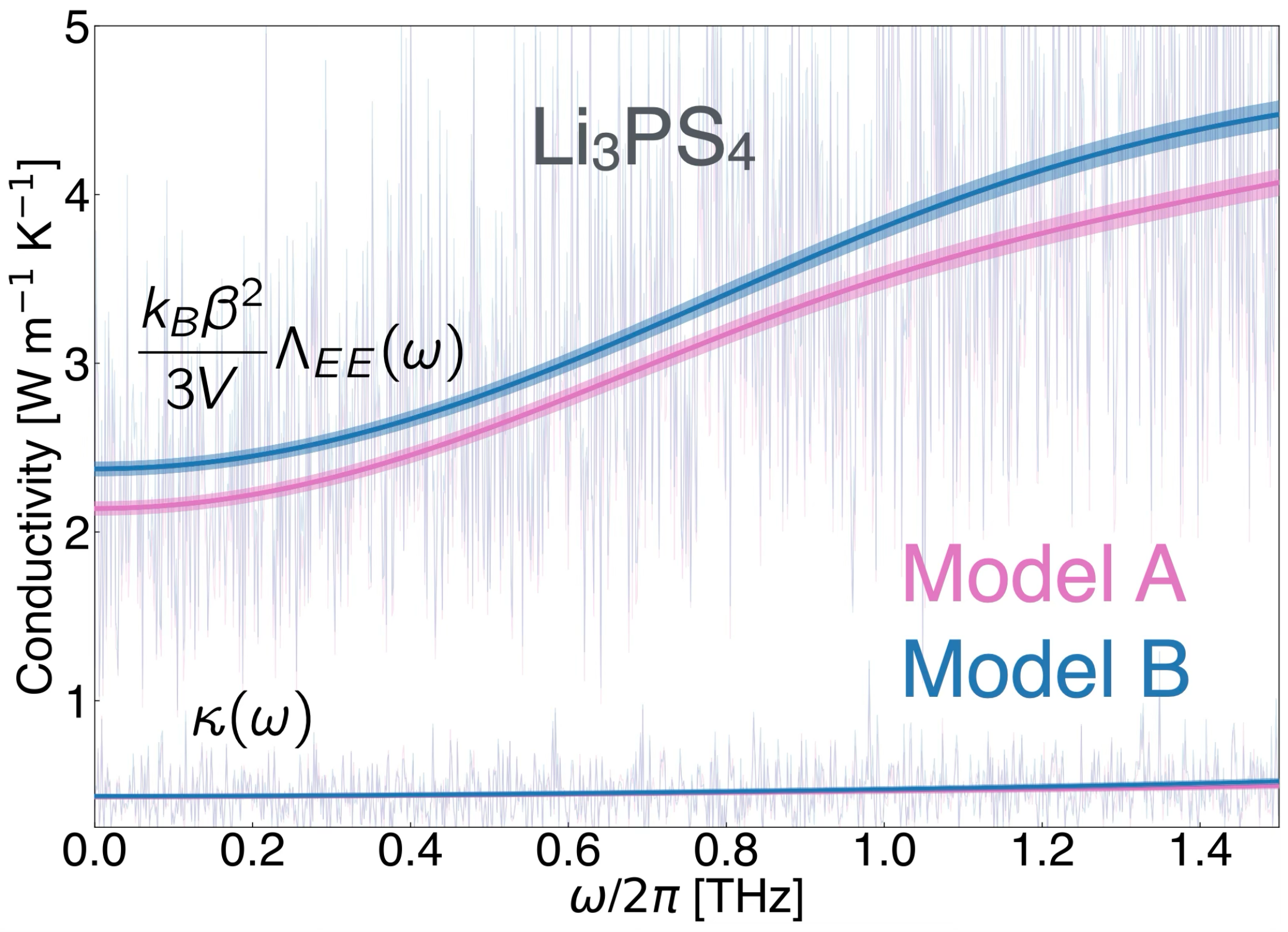}
\caption{\baselineskip=12pt Energy–current power spectra for Li$_3$PS$_4$. Shown are the diagonal element, $\Lambda_{EE}(\omega)$, and the Schur complement of the convective block in the flux cross-spectrum, $\kappa(\omega)$. Their zero-frequency limits give, respectively, the Onsager coefficient $\Lambda_{EE}$ and the thermal conductivity $\kappa$. Different potential models give different $\Lambda_{EE}$ but the same $\kappa$. Data from Ref.~\onlinecite{Tisi2024}, courtesy of the authors.
}
\label{fig:1}
\end{figure}

These considerations have direct consequences for simulations of heat transport with machine-trained neural-network potentials. Redefining atomic energies by adding species-dependent constants leaves the forces unchanged but shifts the energy flux by a convective term, i.e., by a linear combination of mass currents; by convective invariance, the Green–Kubo thermal conductivity is unaffected (see Ref.~\onlinecite{Bertossa2019}). This is illustrated in Fig.~\ref{fig:1}, which reports the power spectrum of the energy current, $\Lambda_{EE}(\omega)$—whose zero-frequency limit yields the Onsager coefficient $\Lambda_{EE}$—for the Li$_3$PS$_4$ solid-state electrolyte, together with its multivariate counterpart, $\kappa(\omega)$ (the \emph{Schur complement of the convective block in the flux cross-spectrum}), whose zero-frequency limit is the thermal conductivity $\kappa$.\cite{Tisi2024} In agreement with the conclusions of this paper and with previous expectations based on the theory of heat transport in multicomponent systems, we find that the $EE$ Onsager coefficient depends on the \emph{energy gauge} (i.e., on the specific local representation of the energy), whereas the thermal conductivity does not; accordingly, $\Lambda_{EE}\neq\kappa$, as it must, because the former is model-dependent whereas the latter is not.

\section*{Acknowledgements}
The considerations presented in this note stem from the questions left open in Ref. \onlinecite{Marcolongo2016}, that have haunted me for more than ten years and were often revisited in conversations with many brilliant young collaborators, too numerous to acknowledge individually. I would nevertheless like to thank Federico Grasselli, Paolo Pegolo, Raffaele Resta, and Davide Tisi for their constructive advice during the writing of this paper and for their critical reading of the text prior to submission.
 I am also grateful to Giovanni Ciccotti, whose ruthless dissection of the hasty presentation of its first version did more for this paper than any number of careful readings, and to Nicola Gigli, for refreshing my rusty elementary calculus. Last but not least, I wish to thank the authors of Ref. \onlinecite{Tisi2024} for providing me with the data reported in Fig. \ref{fig:1}.

This paper is dedicated to my friend and distinguished colleague Annabella Selloni on the occasion of one of her important birthdays.

This work was partially supported by the European Commission through the MaX Centre of Excellence for supercomputing applications (grant number 101093374), by the Italian MUR, through the PRIN project ARES (grant number 2022W2BPCK), and by the Italian National Centre for HPC, Big Data, and Quantum Computing (grant number CN00000013), funded through the Next Generation EU initiative.

\appendix
\section{Derivation of Eqs. (\ref{eq:flux}-\ref{eq:Dl}) and (\ref{eq:J'}-\ref{eq:ff'moment})}
\label{app:J-DI}
To keep the notation as clear as possible--though slightly heavier--in this Appendix I will mostly forgo boldface for vectors and denote their components explicitly with Greek indices.
\subsection{Eqs. (\ref{eq:flux}-\ref{eq:Dl}) }
\begin{gather}
    J_\alpha = \int r_\alpha \dot\epsilon(\rr|\RR,\VV) d\rr \label{eq:J_alpha}\\
    \dot\epsilon(\rr|\RR,\VV) = \sum_{I\beta} \left ( v_{I\beta} \frac{\partial{\epsilon}}{\partial r_{I\beta}} + \frac{f_{I\beta}}{M_I} \frac{\partial\epsilon}{\partial v_{I\beta}}\right ), \label{eq:epsilon_dot}
\end{gather}
where Newton's equations of motion, $\dot\vv_I=\mathbf{f}_I/M_I$, have been used to eliminate accelerations.

Using Eqs. \eqref{eq:energy-density} and \eqref{eq:force-densities}, one has:
\begin{align}
    \begin{aligned}
        \frac{\partial\epsilon}{\partial r_{I\beta}} &= \frac{1}{2}M_I v_I^2 \delta_\beta(\rr-\rr_I) -f_{I\beta}(\rr|\RR) \\
        \frac{\partial\epsilon}{\partial v_{I\beta}} &= M_I v_{I\beta} \delta(\rr-\rr_I),
    \end{aligned} \label{eq:de_epsilon}
\end{align}
where $\delta_\beta(\rr)=\nabla_\beta\delta(\rr)$, $\int \delta_\beta(\rr)g(\rr)d\rr=-\nabla_\beta g(0)$, $\int r_\alpha\delta_\beta(\rr)d\rr = -\delta_{\alpha\beta}$  and $\delta_{\alpha\beta}$ is the Kronecker delta. Substituting Eqs. (\ref{eq:epsilon_dot}-\ref{eq:de_epsilon}) into Eq. \eqref{eq:J_alpha} and using Eq. \eqref{eq:total-forces} gives
\begin{multline}
    J_\alpha = \sum_I v_{I\alpha} \frac{1}{2}M_Iv_I^2 \\ -\sum_{I\beta} v_{I\beta} \left ( \int r_\alpha f_{I\beta}(\rr|\RR) d\rr -r_{I\alpha} f_{I\beta}, \right ), \label{eq:J_final1}
\end{multline}
which can be written as
\begin{multline}
    J_\alpha = \sum_I v_{I\alpha} \frac{1}{2}M_Iv_I^2  \\ -\sum_{I\beta} v_{I\beta} \left ( \int (r_\alpha -r_{I\alpha})f_{I\beta}(\rr|\RR) d\rr \right ), \label{eq:J_final2} 
\end{multline}
thus yielding Eqs. (\ref{eq:flux}-\ref{eq:Dl}).

\subsection{Eqs. (\ref{eq:J'}-\ref{eq:ff'moment})}
The derivation of
Eqs. (\ref{eq:J'}-\ref{eq:ff'moment}) follows similar steps, but applied to the \emph{difference} between two energy densities, $w'=w_2-w_1$. Because $w'$ does not depend on atomic velocities, the expression for the difference betweem the corresponding fluxes has the same structure as Eq. \eqref{eq:J_final2}, except that the kinetic-energy term is missing and the first moment of the force density is taken with respect to $\rr$ rather than $(\rr-\rr_I)$. Since the integral of the force-density differences vanishes, Eq. \eqref{eq:short-sightedness}, this first moment is well-defined, independent of the choice of origin.

\section{A theorem on differential forms in periodic systems}
\label{app:Theorem}
Let $\omega$ be an exact differential form in $\mathbb{R}^N$, periodic with period 1 along each Cartesian axis.
\begin{equation}
    \omega = \mathbf{a}(\mathbf{x}) \cdot d\mathbf{x} = \sum_{n=1}^N a_n(\mathbf{x})\,dx_n,
\end{equation}
where $\mathbf{a}=\{a_1,\dots,a_N\}$ and $\mathbf{x}=\{x_1,\dots,x_N\}$.
Throughout this appendix, vectors in $\mathbb{R}^N$ are denoted by bold letters, and their components in ordinary (italic) math style.
Exactness and periodicity mean
\begin{align}
    \frac{\partial a_n}{\partial x_m} &=\frac{\partial a_m}{\partial x_n}, \label{eq:exactness} \\
    \mathbf{a}(\mathbf{x}+\mathbf{t})&=\mathbf{a}(\mathbf{x})
    \quad\forall\,\mathbf{t}\in\mathbb{Z}^N. \label{eq:periodicity}
\end{align}
The \emph{primitive} of $\omega$, defined by
\begin{equation}
F(\mathbf{x}_0,\mathbf{x}_1)=\int_{\mathbf{x}_0}^{\mathbf{x}_1}\omega,
\end{equation}
only depends on the endpoints of the integration path, because $\omega$ is exact.

\bigskip
\noindent\textbf{Theorem.} 
When the endpoints differ by a lattice vector $\mathbf{t}\in\mathbb{Z}^N$, 
the primitive $F(\mathbf{x}_0,\mathbf{x}_0+\mathbf{t})$ only depends  on $\mathbf{t}$ 
and not on $\mathbf{x}_0$.

\bigskip
\noindent\textbf{Direct proof.} 
Differentiating under the integral sign gives
\begin{equation}
\frac{\partial F}{\partial \mathbf{x}_0}
= \mathbf{a}(\mathbf{x}_0+\mathbf{t}) - \mathbf{a}(\mathbf{x}_0)
= 0,
\end{equation}
because $\mathbf{a}$ is periodic. 
Hence $F(\mathbf{x}_0,\mathbf{x}_0+\mathbf{t})$ is independent of $\mathbf{x}_0$.

\bigskip
\noindent\textbf{Fourier proof.} 
Each component of $\mathbf{a}$ can be expanded in a Fourier series,
\begin{equation}
a_n(\mathbf{x})=\sum_{\mathbf{k}\in\mathbb{Z}^N}
\tilde a_n(\mathbf{k})\,e^{2\pi i\,\mathbf{k}\cdot\mathbf{x}}.
\end{equation}
Since the integral is path-independent, we can choose a straight segment for convenience:
\begin{equation}
F(\mathbf{x}_0,\mathbf{x}_0+\mathbf{t})
=\int_0^1 \bigl ( \mathbf{t}\!\cdot\! \mathbf{a}(\mathbf{x}_0+\lambda\mathbf{t}) \bigr ) d\lambda.
\end{equation}
Substituting the Fourier expansion gives
\begin{multline}
F(\mathbf{x}_0,\mathbf{x}_0+\mathbf{t})
= \\ \sum_{\mathbf{k}}\bigl (\mathbf{t}\!\cdot\!\tilde{\mathbf{a}}(\mathbf{k}) \bigr )
\,e^{2\pi i\,\mathbf{k}\cdot\mathbf{x}_0}
\int_0^1 e^{2\pi i\lambda\,\mathbf{k}\cdot\mathbf{t}}\,d\lambda. \label{eq:F-integral}
\end{multline}
For a any pair of vectors, $\mathbf{t} \text{ and } \mathbf{k} \in\mathbb{Z}^3$, the scalar product $\mathbf{k}\!\cdot\!\mathbf{t}$ is an integer. Therefore, the integral in $\lambda$ in Eq. \eqref{eq:F-integral} vanishes unless $\mathbf{k}\!\cdot\!\mathbf{t}=0$, in which case it equals one.  
Hence
\begin{equation}
    F(\mathbf{x}_0,\mathbf{x}_0+\mathbf{t}) = \sum_{\{\mathbf{k}:\, \mathbf{k} \cdot \mathbf{t}=0\}} (\tilde{\mathbf{a}}(\mathbf{k})\!\cdot\!\mathbf{t})\,e^{2\pi i\,\mathbf{k}\cdot\mathbf{x}_0}. \label{eq:F-sum}
\end{equation}
In reciprocal space, exactness, Eq. \eqref{eq:exactness}, can be expressed as
\begin{equation}
    k_n \tilde a_m(\mathbf{k}) =
    k_m \tilde a_n(\mathbf{k}),
\end{equation}
which means that, for any reciprocal-lattice vector $\mathbf{k}\ne 0$, the Fourier coefficient, $\tilde{\mathbf{a}}_n(\mathbf{k})$, is parallel to $\mathbf{k}$: $\tilde{\mathbf{a}}(\mathbf{k})=\alpha(\mathbf{k})\,\mathbf{k}$.  If $\mathbf{k}\!\cdot\!\mathbf{t}=0$, this implies $ \mathbf{t}\!\cdot\! \tilde{\mathbf{a}}(\mathbf{k}) =0 $. 
We conclude that the only non-vanishing term in the sum, Eq. \eqref{eq:F-sum}, can be the zero-frequency one, 
\begin{equation}
    F(\mathbf{x}_0,\mathbf{x}_0+\mathbf{t}) =\bar{\mathbf{a}}\!\cdot\!\mathbf{t},
\end{equation}
with $\bar{\mathbf{a}} = \tilde{\mathbf{a}}(0)$.

This result shows again that \emph{the integral of a periodic exact form between a point, $\mathbf x_0$, and one of its periodic images, $\mathbf x_0+\mathbf t$, only depends on the difference between the two end points, $\mathbf{t}$.}

\bibliographystyle{apsrev4-2}
\bibliography{cleanedup}

\end{document}